\documentclass[a4paper,aps,prd,10pt,preprintnumbers,showpacs,twocolumn,superscriptaddress,nofootinbib,amsmath,amssymb]{revtex4-1}
\usepackage{graphicx,multirow,cmap,ulem}
\usepackage[utf8]{inputenc}
\usepackage[T1]{fontenc}

\def\K{{\cal K}}

\begin{document}
\title{Quasinormal modes of four-dimensional regular black holes in quasi-topological gravity: Overtones' outburst via WKB method}
\author{R. A. Konoplya}\email{roman.konoplya@gmail.com}
\affiliation{Research Centre for Theoretical Physics and Astrophysics,
Institute of Physics, Silesian University in Opava,
Bezručovo náměstí 13, CZ-74601 Opava, Czech Republic}
\affiliation{Department of Physics, Faculty of Science, University of Hradec Kralove, \\
Rokitanskeho 62/26, Hradec Kralove, 500 03, Czech Republic.}
\begin{abstract}
We study quasinormal modes of scalar, electromagnetic, and Dirac perturbations of four-dimensional regular black holes arising in non-polynomial quasi-topological gravity. Starting from a more general class of metric functions constructed within the same framework, from which two representative cases are selected for detailed analysis, we examine their spectral properties. While the fundamental mode changes smoothly with the regularization parameter, higher overtones display a markedly enhanced sensitivity to near-horizon modifications, leading to the characteristic outburst of overtones. Remarkably, pushing the WKB approximation to sufficiently high orders with Padé resummation already allows one to detect the onset of this effect. Time-domain analysis and the Leaver method confirm that the relative error of the higher-order WKB approach is much smaller than the observed effect. Our results indicate that overtone dynamics provides a sensitive probe of geometrically regular black holes and that high-order WKB methods remain capable of capturing nontrivial spectral features beyond the fundamental mode.
\end{abstract}
\maketitle
%\section{Introduction}

\textbf{Introduction.} Black holes with curvature singularities are a robust prediction of classical general relativity. 
However, the presence of spacetime regions where curvature invariants diverge and geodesic 
evolution becomes incomplete signals a breakdown of the classical theory and strongly suggests 
that a more fundamental description of gravity must modify the near-core geometry. 
This observation has motivated extensive research on {\it regular} black holes, 
in which the central singularity is replaced by a smooth, finite geometry.

Historically, most regular black-hole models were constructed phenomenologically by coupling 
Einstein gravity to non-standard matter sources, such as nonlinear electrodynamics,  
anisotropic or tidal forces in extra dimensional scenarios \cite{Bardeen:1968,Hayward:2005gi,Ansoldi:2008jw,AyonBeato:1998ub,Dymnikova:1992ux,Bronnikov:2000vy,Bronnikov:2024izh,Bronnikov:2005gm,Konoplya:2025ect}. While these models provide valuable insight into possible resolutions of 
singularities, their matter sector is often introduced in an ad hoc manner, and the resulting 
solutions are not always embedded into a fundamental gravitational action \cite{Simpson:2018tsi}.
As a result, an important open question has remained whether regular black holes can arise 
naturally as vacuum solutions of a well-defined gravitational theory.

Recently, it has been shown that regular black holes may emerge in four-dimensional 
purely gravitational theories belonging to the class of non-polynomial quasi-topological 
gravities \cite{Bueno:2025zaj,Borissova:2026wmn}. In this framework, static and spherically symmetric vacuum solutions can be 
obtained in closed form, with the metric functions determined by algebraic relations. 
Remarkably, these solutions are free from curvature singularities and reproduce 
well-known regular geometries --- such as Hayward- or Dymnikova-type metrics --- 
without invoking exotic matter sources. 
Thus, the regularization of the central region is achieved through higher-curvature 
geometric terms alone, providing a conceptually appealing and self-consistent 
realization of non-singular black holes. Previously, polynomial quasi-topological gravity allowed one to obtain various regular black hole solutions in higher than four dimensional spacetimes \cite{Bueno:2024dgm,Bueno:2024eig,Bueno:2025tli,Frolov:2024hhe,Konoplya:2024kih}. In \cite{Konoplya:2024hfg,Konoplya:2025uta,Arbelaez:2026eaz,Arbelaez:2025gwj} it was shown that the truncation of the infinite series containing higher curvature corrections already at the first few order leads to a very good approximation for the observables, such as quasinormal modes, to the final regular solutions for the infinite series.

From the observational point of view, it is crucial to understand how such geometrically regular black holes respond to perturbations and how their strong-field properties differ from those of the Schwarzschild solution. The quasinormal mode (QNM) spectrum governs the ringdown phase of gravitational-wave signals \cite{Kokkotas:1999bd,Konoplya:2011qq,Berti:2009kk,Bolokhov:2025rng}.

In this work, starting from generation of some generic black hole solution, we then study the quasinormal modes of two particular models of regular black holes arising from four-dimensional non-polynomial quasi-topological gravity. 
By analyzing scalar, electromagnetic, and Dirac perturbations, 
we determine how the absence of a singular core and the presence of higher-curvature corrections modify the ringdown spectrum. 

An interesting observation concerns the behavior of the first few overtones. As shown in \cite{Konoplya:2022pbc}, the first few overtones can be particularly sensitive to the near-horizon portion of the geometry, effectively acting as a ``sound'' of the event horizon \cite{Konoplya:2023hqb}. This effect---often termed the {\it outburst of overtones}---has been reported for a number of black-hole metrics that deviate appreciably from the Schwarzschild/Kerr spacetimes only in a narrow region close to the horizon \cite{Konoplya:2022pbc,Konoplya:2023hqb}. In contrast, the fundamental mode is typically controlled mainly by the geometry around the peak of the effective potential and therefore may respond much more weakly to deformations confined to the near-horizon layer. The overtones, on the other hand, can react to even small changes of the background in that region. 

Because of this enhanced sensitivity, extracting the first several overtones is often difficult with fast ``black-box'' techniques such as standard-order WKB or time-domain fitting, and one usually resorts to more sophisticated approaches, e.g., Frobenius-type expansions or pseudospectral methods. However, in the present case we demonstrate that pushing the WKB approximation to the 14th--16th orders and applying Pad\'e resummation, as advocated in \cite{Matyjasek:2019eeu} already allows one to access several higher overtones (including modes with $n>\ell$) and to observe the onset of the overtone outburst even within the WKB framework.

%\section{Perturbation equations and effective potentials}\label{sec:wavelike}

\textbf{Perturbation equations and effective potentials.} 
Before specifying the metric ansatz, let us briefly outline the theoretical framework of non-polynomial quasi-topological gravity (NP-QTG) discussed in \cite{Borissova:2026wmn}. Quasi-topological gravities belong to a class of higher-curvature theories constructed so that, despite containing nontrivial curvature invariants beyond the Einstein–Hilbert term, they admit second-order field equations for static and spherically symmetric spacetimes. In four dimensions, purely polynomial quasi-topological densities are either topological or trivial; however, non-polynomial extensions allow one to construct genuinely dynamical higher-curvature corrections while preserving the remarkable property that the field equations reduce to an algebraic relation for the metric function under spherical symmetry.

The action of NP-QTG can be written schematically as
\begin{equation}
S = \frac{1}{16\pi G}\int d^4 x \sqrt{-g}
\left[
R + \mathcal{F}(\mathcal{R})
\right],
\end{equation}
where $\mathcal{R}$ denotes a specific higher-curvature scalar constructed from contractions of the Riemann tensor, and $\mathcal{F}$ is a non-polynomial function chosen such that the theory remains ghost-free around maximally symmetric backgrounds and yields second-order equations for static spherical configurations. The precise functional form of $\mathcal{F}$ is engineered so that the trace of the field equations collapses to a total derivative in the spherically symmetric sector.

As a consequence, for the metric ansatz
\begin{equation}
ds^2 = -f(r) dt^2 + \frac{dr^2}{f(r)} + r^2 d\Omega^2,
\end{equation}
after introducing the parametrization,
\begin{equation}
f(r)=1-r^2\psi(r),
\end{equation}
the full set of modified gravitational equations reduces to a single generating equation of the form  \cite{Borissova:2026wmn}
\begin{equation}\label{generatingEQ}
\frac{d}{dr}\left(r^3 h(\psi)\right)=0.
\end{equation}

Here the function $h(\psi)$ encodes the specific higher-curvature structure of the theory and is determined by the choice of the non-polynomial function $\mathcal{F}$. This remarkable reduction allows one to obtain static vacuum solutions in closed form without introducing exotic matter sources.

In particular, suitable choices of $h(\psi)$ lead to asymptotically flat black-hole geometries that are completely regular at the center. The corresponding function $h(\psi)$ is defined implicitly as the inverse of the relation
\begin{equation}
\psi = \frac{h}{\left(1 + \alpha^{\,\nu} h^{\nu}\right)^{\mu/3\nu}},
\end{equation}
which admits a power-series expansion in $h$. As shown in \cite{Tsuda:2026xjc}, the corresponding inverse series exists, converges, and its coefficients can be computed explicitly for arbitrary positive integers $\mu$ and $\nu$.

Following the procedure developed in \cite{Tsuda:2026xjc} for higher-dimensional black holes, one finds that the metric function takes the generic form
\begin{equation}\label{solution}
f(r)=1-
\frac{2 M r^{\mu-1}}
{\left(r^{\nu}+\alpha^{\,\nu/3} (2 M)^{\nu/3}\right)^{\mu/\nu}},
\end{equation}
where $M$ appears as the integration constant of Eq.~\eqref{generatingEQ} and is identified with the asymptotic mass. The parameter $\alpha$ has dimensions of length squared; accordingly, we write $\alpha = l^2$. Particular models discussed in \cite{Borissova:2026wmn} are recovered for specific choices of the integers $\mu$ and $\nu$.

We analyze two specific realizations of the above generic metric function $f(r)$ \cite{Borissova:2026wmn}:
\[
\begin{array}{rclll}
f(r) &=& 1 - \dfrac{2 M r^2}{\sqrt{4 l^4 M^2 + r^6}}, &
\quad \textit{Model I} & (\mu=3, \nu=6),\\
f(r) &=& 1 - \dfrac{2 M r^2}{2 Ml^2 + r^3}, 
& \quad \textit{Model II} & (\mu=3, \nu=3).
\end{array}
\]
Both metrics arise as four–dimensional solutions within quasi-topological gravity \cite{Borissova:2026wmn}. In what follows, all dimensional quantities are expressed in units of the mass by setting $M=1$.

The second metric function coincides, after a simple redefinition of parameters, with the Hayward regular black hole solution \cite{Hayward:2005gi}. Since its quasinormal spectrum has already been studied in detail and reviewed in \cite{Konoplya:2022hll,Konoplya:2023ppx,Malik:2024tuf,Bolokhov:2025egl,Liang:2025rbe,Ahmed:2026eew,Malik:2025dxn,Lin:2013ofa,DuttaRoy:2022ytr,Pedraza:2021hzw,Spina:2025wxb}, we will not repeat that analysis here. Instead, Model II will serve as a useful auxiliary background illustrating the ability of high-order WKB techniques to reveal nontrivial behavior of higher overtones.

\medskip

We consider test perturbations of scalar, electromagnetic, and Dirac fields propagating in the black hole background. Their covariant field equations are
\begin{subequations}\label{coveqs}
\begin{eqnarray}
\frac{1}{\sqrt{-g}}\partial_\mu\!\left(\sqrt{-g}\, g^{\mu\nu}\partial_\nu \Phi \right) &=& 0,
\\
\frac{1}{\sqrt{-g}}\partial_\mu\!\left(\sqrt{-g}\, F_{\rho\sigma} g^{\rho\nu} g^{\sigma\mu} \right) &=& 0,
\\
\gamma^\alpha \left(\partial_\alpha - \Gamma_\alpha \right)\Upsilon &=& 0,
\end{eqnarray}
\end{subequations}
where $\Phi$ is the scalar field, $A_\mu$ is the electromagnetic four-potential, $F_{\mu\nu}=\partial_\mu A_\nu-\partial_\nu A_\mu$ is the electromagnetic tensor, $\Upsilon$ denotes the Dirac spinor, $\gamma^\alpha$ are the curved-space gamma matrices, and $\Gamma_\alpha$ are the spin connections defined in the tetrad formalism.

After separating variables and decomposing the angular dependence in spherical harmonics, the radial part of each perturbation equation reduces to a Schr\"odinger-type wave equation~\cite{Kokkotas:1999bd,Berti:2009kk,Konoplya:2011qq}:
\begin{equation}\label{wave-equation}
\frac{d^2\Psi}{dr_*^2} + \left(\omega^2 - V(r)\right)\Psi = 0,
\end{equation}
where $\omega$ is the (generally complex) frequency. The tortoise coordinate $r_*$ is defined by
\begin{equation}\label{tortoise}
\frac{dr_*}{dr} = \frac{1}{f(r)},
\end{equation}
which maps the event horizon to $r_* \to -\infty$ and spatial infinity to $r_* \to +\infty$.

For bosonic fields with spin $s=0$ (scalar) and $s=1$ (electromagnetic), the effective potential can be written in the unified form
\begin{equation}\label{potentialScalar}
V(r) = f(r)\frac{\ell(\ell+1)}{r^2}
+ \frac{1-s}{r}\,\frac{d^2 r}{dr_*^2},
\end{equation}
where $\ell = s, s+1, s+2, \ldots$ is the multipole number. The second term contributes only in the scalar case.

For the Dirac field ($s=1/2$), the radial equations lead to two supersymmetric partner potentials,
\begin{equation}
V_\pm(r) = W^2 \pm \frac{dW}{dr_*},
\qquad 
W = \left(\ell+\frac{1}{2}\right)\frac{\sqrt{f(r)}}{r}.
\end{equation}
These potentials are isospectral, i.e., they share identical quasinormal spectra. The corresponding wave functions are related by the Darboux transformation
\begin{equation}\label{psi}
\Psi_+ \propto \left(W + \frac{d}{dr_*}\right)\Psi_-.
\end{equation}
Therefore, it is sufficient to compute the quasinormal frequencies for only one of them. In the following analysis, we concentrate on $V_+(r)$, since its structure is more favorable for applying the WKB approximation, especially at higher orders.

\begin{figure*}
\resizebox{\linewidth}{!}{\includegraphics{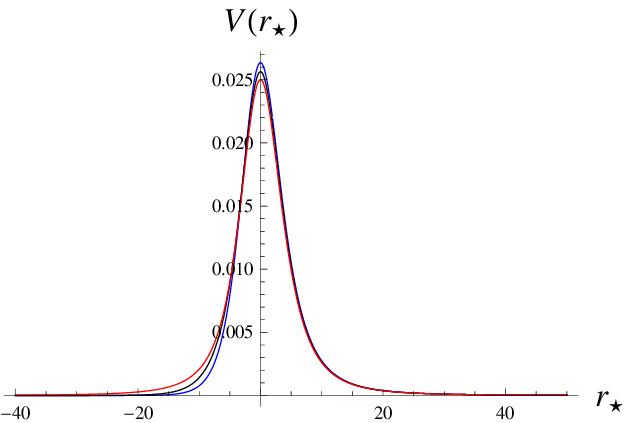}\includegraphics{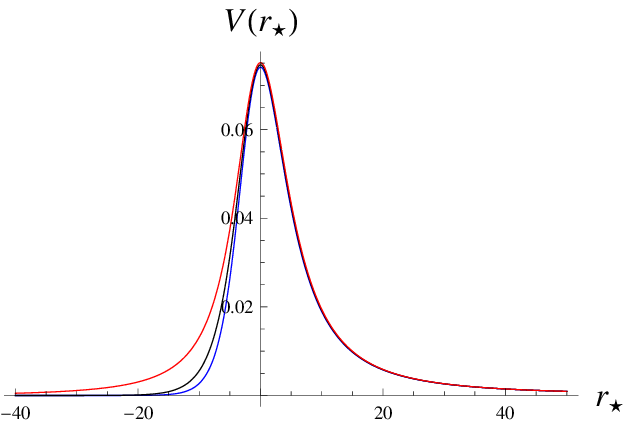}}
\caption{Effective potential as a function of the tortoise coordinate $r^{*}$ for $\ell=0$ scalar perturbations and  $\ell=1$ electromagnetic perturbations: $M=1$; $h=0.1$ (blue), $h=5$ (black) and $h=9.4$ (red).}\label{fig:potL2up2}
\end{figure*}

%\section{WKB formula}\label{sec:WKB}

\begin{figure}
\resizebox{\linewidth}{!}{\includegraphics{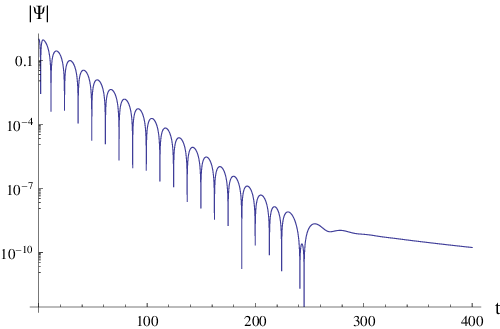}}
\caption{Time domain profile for electromagnetic perturbations of the black hole model I. Here we have $M=1$, $\ell=1$, $h=4 l^4=9.4$. The time-domain integration gives the fundamental mode $\omega = 0.249667 - 0.0827673 i$, while the WKB method $\omega = 0.249666 - 0.082771 i$.}\label{fig:TD}
\end{figure}

\begin{table*}
\centering
\begin{tabular}{c|ccc|ccc}
\hline
& \multicolumn{3}{c|}{$\ell=0$} & \multicolumn{3}{c}{$\ell=1$} \\
$4 l^4$ 
& WKB-16 ($m=8$) 
& WKB-14 ($m=7$) 
& diff (\%) 
& WKB-16 ($m=8$) 
& WKB-14 ($m=7$) 
& diff (\%) \\
\hline
1.0 & $0.110681-0.103666 i$ & $0.110723-0.103617 i$ & 0.0420 
    & $0.293038-0.096869 i$ & $0.293038-0.096869 i$ & 0 \\
3.0 & $0.110995-0.101114 i$ & $0.110686-0.101346 i$ & 0.257 
    & $0.293114-0.095212 i$ & $0.293116-0.095214 i$ & 0.0007 \\
5.0 & $0.110693-0.098361 i$ & $0.110865-0.097933 i$ & 0.311 
    & $0.292966-0.093489 i$ & $0.292966-0.093488 i$ & 0.0002 \\
7.0 & $0.108994-0.094642 i$ & $0.108779-0.094268 i$ & 0.299 
    & $0.292536-0.091800 i$ & $0.292535-0.091802 i$ & 0.001 \\
9.0 & $0.105514-0.093851 i$ & $0.105561-0.093833 i$ & 0.0353 
    & $0.291881-0.090295 i$ & $0.291881-0.090295 i$ & 0.0002 \\
9.4 & $0.105188-0.093854 i$ & $0.105228-0.093839 i$ & 0.0307 
    & $0.291737-0.090020 i$ & $0.291738-0.090021 i$ & 0.0004 \\
\hline
\end{tabular}
\caption{Fundamental quasinormal modes ($n=0$) of scalar perturbations for the regular black hole I ($M=1$) calculated using the WKB formula at different orders with Pad\'e approximants. Results are shown for $\ell=0$ and $\ell=1$. The last columns display the relative difference between the 16th- and 14th-order WKB results (in percent).}
\end{table*}

\begin{table*}
\centering
\begin{tabular}{c|ccc|ccc}
\hline
& \multicolumn{3}{c|}{Electromagnetic ($\ell=1$)} 
& \multicolumn{3}{c}{Dirac ($\ell=1/2$)} \\
$4 l^4$ 
& WKB-16 ($m=8$) 
& WKB-14 ($m=7$) 
& diff (\%) 
& WKB-16 ($m=8$) 
& WKB-14 ($m=7$) 
& diff (\%) \\
\hline
1.0 & $0.248684-0.091625 i$ & $0.248685-0.091626 i$ & 0.0005 
    & $0.182759-0.095926 i$ & $0.182705-0.095924 i$ & 0.0261 \\
3.0 & $0.249427-0.089741 i$ & $0.249427-0.089742 i$ & 0  
    & $0.182073-0.093708 i$ & $0.182077-0.093706 i$ & 0.0022 \\
5.0 & $0.249946-0.087625 i$ & $0.249945-0.087627 i$ & 0.0007 
    & $0.180916-0.091403 i$ & $0.180924-0.091405 i$ & 0.0039 \\
7.0 & $0.250084-0.085360 i$ & $0.250083-0.085360 i$ & 0.0003 
    & $0.179051-0.089316 i$ & $0.179051-0.089316 i$ & 0.0002 \\
9.0 & $0.249766-0.083173 i$ & $0.249765-0.083173 i$ & 0.0004 
    & $0.176829-0.088181 i$ & $0.176828-0.088182 i$ & 0.0007 \\
9.4 & $0.249666-0.082771 i$ & $0.249664-0.082771 i$ & 0.0004 
    & $0.176446-0.088034 i$ & $0.176445-0.088034 i$ & 0.0010 \\
\hline
\end{tabular}

%\textbf{What is the difference between 0 and $<10^{-4}$ ?}
\caption{Fundamental quasinormal modes ($n=0$) of electromagnetic ($\ell=1$) and Dirac ($\ell=1/2$) perturbations for the regular black hole I ($M=1$), computed using 16th- and 14th-order WKB formulas with Pad\'e approximants. The last columns show the relative difference between the two WKB orders (in percent).}
\end{table*}

\begin{table*}
\centering
\begin{tabular}{c|ccc|ccc}
\hline
& \multicolumn{3}{c|}{$\ell=1$, $n=1$} 
& \multicolumn{3}{c}{$\ell=1$, $n=2$} \\
$4 l^4$ 
& WKB-16 ($m=8$) 
& WKB-14 ($m=7$) 
& diff (\%) 
& WKB-16 ($m=8$) 
& WKB-14 ($m=7$) 
& diff (\%) \\
\hline
1.0 & $0.215060-0.290248 i$ & $0.215216-0.290330 i$ & 0.0489
    & $0.172278-0.516055 i$ & $0.175355-0.517576 i$ & 0.631 \\
1.4 & $0.215232-0.288722 i$ & $0.215334-0.288723 i$ & 0.0281
    & $0.172208-0.512573 i$ & $0.174909-0.514262 i$ & 0.589 \\
1.8 & $0.215311-0.287124 i$ & $0.215358-0.287107 i$ & 0.0140
    & $0.171921-0.509091 i$ & $0.173973-0.510727 i$ & 0.488 \\
2.2 & $0.215296-0.285491 i$ & $0.215314-0.285478 i$ & 0.0062
    & $0.171226-0.505618 i$ & $0.172531-0.507026 i$ & 0.360 \\
2.6 & $0.215200-0.283834 i$ & $0.215205-0.283828 i$ & 0.0022
    & $0.169888-0.502146 i$ & $0.170596-0.503239 i$ & 0.246 \\
3.0 & $0.215027-0.282152 i$ & $0.215028-0.282151 i$ & 0.0004
    & $0.167830-0.498628 i$ & $0.168203-0.499474 i$ & 0.176 \\
3.4 & $0.214771-0.280443 i$ & $0.214771-0.280443 i$ & 0
    & $0.165166-0.495115 i$ & $0.165397-0.495874 i$ & 0.152 \\
3.8 & $0.214421-0.278706 i$ & $0.214421-0.278705 i$ & 0.0005
    & $0.162029-0.491762 i$ & $0.162225-0.492602 i$ & 0.167 \\
4.2 & $0.213962-0.276951 i$ & $0.213964-0.276947 i$ & 0.0013
    & $0.158509-0.488741 i$ & $0.158746-0.489801 i$ & 0.212 \\
4.6 & $0.213386-0.275195 i$ & $0.213389-0.275190 i$ & 0.0017
    & $0.154672-0.486194 i$ & $0.155048-0.487553 i$ & 0.276 \\
5.0 & $0.212685-0.273467 i$ & $0.212688-0.273463 i$ & 0.0014
    & $0.150600-0.484229 i$ & $0.151240-0.485842 i$ & 0.342 \\
5.4 & $0.211853-0.271793 i$ & $0.211855-0.271791 i$ & 0.0008
    & $0.146393-0.482902 i$ & $0.147398-0.484559 i$ & 0.384 \\
5.8 & $0.210889-0.270200 i$ & $0.210889-0.270199 i$ & 0.0003
    & $0.142147-0.482208 i$ & $0.143481-0.483557 i$ & 0.377 \\
6.2 & $0.209799-0.268710 i$ & $0.209799-0.268710 i$ & $<10^{-4}$
    & $0.137886-0.482113 i$ & $0.139254-0.482806 i$ & 0.306 \\
6.6 & $0.208604-0.267351 i$ & $0.208604-0.267351 i$ & 0
    & $0.133512-0.482707 i$ & $0.134360-0.482715 i$ & 0.169 \\
7.0 & $0.207336-0.266148 i$ & $0.207336-0.266148 i$ & 0
    & $0.129045-0.484590 i$ & $0.129129-0.484523 i$ & 0.0213 \\
7.4 & $0.206041-0.265112 i$ & $0.206041-0.265112 i$ & 0
    & $0.126150-0.488728 i$ & $0.126116-0.488426 i$ & 0.0603 \\
7.8 & $0.204769-0.264232 i$ & $0.204769-0.264231 i$ & $<10^{-4}$
    & $0.127166-0.492083 i$ & $0.125811-0.491115 i$ & 0.328 \\
8.2 & $0.203564-0.263465 i$ & $0.203564-0.263464 i$ & 0.0002
    & $0.128156-0.491190 i$ & $0.125507-0.491604 i$ & 0.528 \\
8.6 & $0.202470-0.262744 i$ & $0.202470-0.262744 i$ & 0.0001
    & $0.126505-0.490044 i$ & $0.124222-0.491569 i$ & 0.542 \\
9.0 & $0.201486-0.262041 i$ & $0.201486-0.262041 i$ & 0.0002
    & $0.124108-0.489907 i$ & $0.122408-0.491688 i$ & 0.487 \\
9.4 & $0.200478-0.261367 i$ & $0.200481-0.261364 i$ & 0.0012
    & $0.122027-0.490230 i$ & $0.120527-0.492017 i$ & 0.462 \\
\hline
\end{tabular}
\caption{Quasinormal modes of electromagnetic perturbations ($\ell=1$) for the regular black hole I ($M=1$). Left block: first overtone ($n=1$). Right block: second overtone ($n=2$). Frequencies are calculated using 16th- and 14th-order WKB formulas with Pad\'e approximants. The last columns show the relative difference between the two WKB orders (in percent).}
\end{table*}

\begin{table}
\begin{tabular}{c c c c}
\hline
$4 l^4$ & WKB16 ($m=8$) & WKB14 ($m=7$) & diff. ($\%$)  \\
\hline
$1.$ & $0.142932-0.766977 i$ & $0.141564-0.752031 i$ & $1.92\%$\\
$1.4$ & $0.139164-0.761494 i$ & $0.138049-0.745454 i$ & $2.08\%$\\
$1.8$ & $0.133814-0.755019 i$ & $0.133672-0.739608 i$ & $2.01\%$\\
$2.2$ & $0.126393-0.748847 i$ & $0.128288-0.734266 i$ & $1.94\%$\\
$2.6$ & $0.115603-0.743194 i$ & $0.121923-0.729484 i$ & $2.01\%$\\
$3.$ & $0.101927-0.734563 i$ & $0.114776-0.725494 i$ & $2.12\%$\\
$3.4$ & $0.092199-0.724533 i$ & $0.107151-0.722672 i$ & $2.06\%$\\
$3.8$ & $0.083831-0.719408 i$ & $0.099383-0.721461 i$ & $2.17\%$\\
$4.2$ & $0.073790-0.718236 i$ & $0.091770-0.722250 i$ & $2.55\%$\\
$4.6$ & $0.062498-0.720670 i$ & $0.084557-0.725268 i$ & $3.11\%$\\
$5.$ & $0.051467-0.727470 i$ & $0.077931-0.730579 i$ & $3.65\%$\\
$5.4$ & $0.043332-0.739090 i$ & $0.072006-0.738213 i$ & $3.87\%$\\
$5.8$ & $0.041156-0.754435 i$ & $0.066850-0.748445 i$ & $3.49\%$\\
$6.2$ & $0.046659-0.771986 i$ & $0.062672-0.762317 i$ & $2.42\%$\\
$6.6$ & $0.062425-0.786639 i$ & $0.060640-0.782755 i$ & $0.542\%$\\
$7.$ & $0.065049-0.816714 i$ & $0.069135-0.817604 i$ & $0.510\%$\\
\hline
\end{tabular}
\caption{The third overtone ($n=3$) of the $\ell=1$ electromagnetic perturbations for black hole model I ($M=1$) is computed using the WKB approximation at different orders with Pad\'e approximants. The last columns display the relative difference between the two WKB orders. Although the WKB method is only marginally accurate in this case, the estimated relative error (a few percents) remains significantly smaller than the deviation observed in the real part of the frequency (tens to a hundred percents), where the characteristic ``outburst of overtones'' typically occurs. At $h>7$ the accuracy of the WKB method become insufficient even at high orders.}
\end{table}

\begin{table*}
%\begin{scriptsize}
\centering
\scalebox{0.8}[1.0]{
\begin{tabular}{c|ccc|ccc|ccc}
\hline
& \multicolumn{3}{c|}{$\ell=1$, $n=1$} 
& \multicolumn{3}{c|}{$\ell=1$, $n=2$} 
& \multicolumn{3}{c}{$\ell=1$, $n=3$} \\
$4 l^4$ 
& WKB-16 ($m=8$)
& WKB-14 ($m=7$)
& diff (\%) 
& WKB-16 ($m=8$) 
& WKB-14 ($m=7$)
& diff (\%) 
& WKB-16 ($m=8$)
& WKB-14 ($m=7$)
& diff (\%) \\
\hline
1.0 & $0.263920-0.303052 i$ & $0.264113-0.303022 i$ & 0.049
    & $0.225764-0.533511 i$ & $0.226453-0.532590 i$ & 0.198
    & $0.195687-0.778289 i$ & $0.195435-0.779062 i$ & 0.101 \\
1.4 & $0.263591-0.301732 i$ & $0.263777-0.301720 i$ & 0.047
    & $0.224452-0.530865 i$ & $0.224533-0.529960 i$ & 0.158
    & $0.190577-0.772073 i$ & $0.190548-0.772159 i$ & 0.011 \\
1.8 & $0.263141-0.300318 i$ & $0.263406-0.300398 i$ & 0.070
    & $0.223362-0.527750 i$ & $0.223232-0.527766 i$ & 0.023
    & $0.189535-0.764874 i$ & $0.183949-0.766268 i$ & 0.731 \\
2.2 & $0.262882-0.298812 i$ & $0.262984-0.299091 i$ & 0.075
    & $0.221190-0.524126 i$ & $0.221186-0.524212 i$ & 0.015
    & $0.177514-0.764487 i$ & $0.176650-0.762225 i$ & 0.309 \\
2.6 & $0.262484-0.297554 i$ & $0.262592-0.297937 i$ & 0.100
    & $0.218318-0.521116 i$ & $0.218407-0.521301 i$ & 0.036
    & $0.168955-0.760994 i$ & $0.169147-0.760392 i$ & 0.081 \\
3.0 & $0.261939-0.296260 i$ & $0.261969-0.296180 i$ & 0.022
    & $0.215162-0.518486 i$ & $0.215358-0.518667 i$ & 0.048
    & $0.160026-0.761597 i$ & $0.160488-0.761422 i$ & 0.064 \\
3.4 & $0.261313-0.294956 i$ & $0.261315-0.294948 i$ & 0.002
    & $0.211756-0.516217 i$ & $0.212091-0.516421 i$ & 0.070
    & $0.149621-0.763707 i$ & $0.151000-0.763990 i$ & 0.181 \\
3.8 & $0.260608-0.293661 i$ & $0.260608-0.293661 i$ & 0
    & $0.208084-0.514202 i$ & $0.208642-0.514881 i$ & 0.158
    & $0.140437-0.767382 i$ & $0.143323-0.768187 i$ & 0.384 \\
4.2 & $0.259817-0.292392 i$ & $0.259818-0.292389 i$ & 0.001
    & $0.204051-0.512716 i$ & $0.205067-0.514072 i$ & 0.307
    & $0.134191-0.773205 i$ & $0.138437-0.773466 i$ & 0.542 \\
4.6 & $0.258932-0.291175 i$ & $0.258946-0.291170 i$ & 0.004
    & $0.199810-0.511975 i$ & $0.201678-0.513769 i$ & 0.471
    & $0.131694-0.779492 i$ & $0.135834-0.778352 i$ & 0.543 \\
5.0 & $0.257922-0.290000 i$ & $0.257995-0.290015 i$ & 0.019
    & $0.195719-0.512109 i$ & $0.198684-0.513699 i$ & 0.614
    & $0.131598-0.784131 i$ & $0.134176-0.782250 i$ & 0.401 \\
5.4 & $0.256917-0.288791 i$ & $0.256976-0.288931 i$ & 0.040
    & $0.192215-0.512918 i$ & $0.196038-0.513569 i$ & 0.708
    & $0.131497-0.787157 i$ & $0.132393-0.785635 i$ & 0.221 \\
5.8 & $0.255880-0.287815 i$ & $0.255894-0.287936 i$ & 0.032
    & $0.189465-0.513902 i$ & $0.193445-0.513119 i$ & 0.741
    & $0.129967-0.790934 i$ & $0.129793-0.790094 i$ & 0.107 \\
6.2 & $0.254748-0.286955 i$ & $0.254769-0.287051 i$ & 0.026
    & $0.187274-0.514668 i$ & $0.189957-0.512330 i$ & 0.650
    & $0.132559-0.800255 i$ & $0.130297-0.801372 i$ & 0.311 \\
6.6 & $0.253600-0.286217 i$ & $0.253629-0.286287 i$ & 0.020
    & $0.185414-0.515283 i$ & $0.185242-0.514174 i$ & 0.205
    & $0.139902-0.811776 i$ & $0.122554-0.826874 i$ & 2.79 \\
7.0 & $0.252483-0.285595 i$ & $0.252510-0.285638 i$ & 0.013
    & $0.184040-0.516360 i$ & $0.184040-0.516360 i$ & 0
    & $0.094986-0.833889 i$ & $0.098134-0.805464 i$ & 3.41 \\
\hline
\end{tabular}}

%\textbf{It seems this is the only table where m is not specified.\\ You can use %$\backslash resizebox\{\backslash linewidth\}\{!\}\{...\}$ here}
\caption{Quasinormal modes of scalar perturbations ($\ell=1$) for the regular black hole I ($M=1$). Results are shown for the first three overtones ($n=1,2,3$), computed using 16th- and 14th-order WKB formulas with Pad\'e approximants. The last columns give the relative difference between the two WKB orders (in percent).}
%\end{scriptsize}
\end{table*}

\begin{table*}
\centering
\scalebox{0.75}[1.0]{
\renewcommand{\arraystretch}{1.05}
\setlength{\tabcolsep}{5.2pt}
\begin{tabular}{c|c c c|c c c|c c c}
\hline
\multicolumn{1}{c|}{\multirow{2}{*}{$2 M l^2$}} &
\multicolumn{3}{c|}{$\ell=1$, scalar, $n=0$} &
\multicolumn{3}{c|}{$\ell=1$, scalar, $n=1$} &
\multicolumn{3}{c}{$\ell=1$, scalar, $n=2$} \\
\cline{2-10}
& WKB16 ($m=8$) & WKB14 ($m=7$) & diff. &
  WKB16 ($m=8$) & WKB14 ($m=7$) & diff. &
  WKB16 ($m=8$) & WKB14 ($m=7$) & diff. \\
\hline
$0.02$ & $0.248565-0.092346 i$ & $0.248565-0.092346 i$ & $0.000$ &
        $0.215015-0.293111 i$ & $0.215004-0.293109 i$ & $0.003$ &
        $0.175050-0.524015 i$ & $0.174952-0.524171 i$ & $0.033$ \\
$0.10$ & $0.249789-0.091754 i$ & $0.249789-0.091754 i$ & $0.000$ &
        $0.217050-0.290772 i$ & $0.217046-0.290755 i$ & $0.005$ &
        $0.177683-0.518791 i$ & $0.177401-0.519173 i$ & $0.087$ \\
$0.20$ & $0.251361-0.090951 i$ & $0.251361-0.090951 i$ & $0.000$ &
        $0.219615-0.287608 i$ & $0.219620-0.287575 i$ & $0.009$ &
        $0.180932-0.512030 i$ & $0.180756-0.512254 i$ & $0.052$ \\
$0.30$ & $0.252981-0.090067 i$ & $0.252980-0.090067 i$ & $0.000$ &
        $0.222192-0.284186 i$ & $0.222192-0.284187 i$ & $0.000$ &
        $0.183812-0.505277 i$ & $0.183760-0.505276 i$ & $0.010$ \\
$0.40$ & $0.254647-0.089091 i$ & $0.254647-0.089091 i$ & $0.000$ &
        $0.224754-0.280364 i$ & $0.224922-0.280559 i$ & $0.072$ &
        $0.185890-0.497138 i$ & $0.186036-0.497387 i$ & $0.054$ \\
$0.50$ & $0.256361-0.088007 i$ & $0.256361-0.088007 i$ & $0.000$ &
        $0.227165-0.276138 i$ & $0.227401-0.275977 i$ & $0.080$ &
        $0.187965-0.487699 i$ & $0.187898-0.487902 i$ & $0.041$ \\
$0.60$ & $0.258118-0.086797 i$ & $0.258118-0.086797 i$ & $0.000$ &
        $0.229479-0.271457 i$ & $0.229536-0.271341 i$ & $0.036$ &
        $0.189559-0.477205 i$ & $0.189550-0.477209 i$ & $0.002$ \\
$0.70$ & $0.259911-0.085437 i$ & $0.259911-0.085437 i$ & $0.000$ &
        $0.231537-0.266209 i$ & $0.231527-0.266209 i$ & $0.003$ &
        $0.189921-0.466087 i$ & $0.189876-0.466066 i$ & $0.010$ \\
$0.80$ & $0.261725-0.083900 i$ & $0.261726-0.083900 i$ & $0.000$ &
        $0.233060-0.260370 i$ & $0.233060-0.260385 i$ & $0.004$ &
        $0.187575-0.454906 i$ & $0.187518-0.454552 i$ & $0.073$ \\
$0.90$ & $0.263533-0.082153 i$ & $0.263533-0.082153 i$ & $0.000$ &
        $0.233809-0.254009 i$ & $0.233812-0.254020 i$ & $0.003$ &
        $0.181044-0.443484 i$ & $0.182047-0.443685 i$ & $0.214$ \\
$0.92$ & $0.263890-0.081776 i$ & $0.263890-0.081776 i$ & $0.000$ &
        $0.233827-0.252686 i$ & $0.233836-0.252700 i$ & $0.005$ &
        $0.179447-0.441547 i$ & $0.180644-0.441810 i$ & $0.257$ \\
$0.94$ & $0.264244-0.081390 i$ & $0.264244-0.081390 i$ & $0.000$ &
        $0.233791-0.251364 i$ & $0.233810-0.251372 i$ & $0.006$ &
        $0.177742-0.439797 i$ & $0.179193-0.440069 i$ & $0.311$ \\
$0.96$ & $0.264595-0.080993 i$ & $0.264595-0.080993 i$ & $0.000$ &
        $0.233714-0.250043 i$ & $0.233731-0.250041 i$ & $0.005$ &
        $0.175976-0.438257 i$ & $0.177734-0.438427 i$ & $0.374$ \\
$0.98$ & $0.264942-0.080586 i$ & $0.264942-0.080586 i$ & $0.000$ &
        $0.233586-0.248718 i$ & $0.233596-0.248713 i$ & $0.003$ &
        $0.174200-0.436937 i$ & $0.176294-0.436759 i$ & $0.447$ \\
$1.00$ & $0.265285-0.080170 i$ & $0.265285-0.080169 i$ & $0.000$ &
        $0.233400-0.247398 i$ & $0.233404-0.247394 i$ & $0.002$ &
        $0.172439-0.435830 i$ & $0.174689-0.434640 i$ & $0.543$ \\
$1.02$ & $0.265622-0.079743 i$ & $0.265622-0.079743 i$ & $0.000$ &
        $0.233154-0.246093 i$ & $0.233154-0.246093 i$ & $0.000$ &
        $0.170634-0.434950 i$ & $0.170158-0.432205 i$ & $0.596$ \\
$1.04$ & $0.265953-0.079307 i$ & $0.265953-0.079307 i$ & $0.000$ &
        $0.232850-0.244819 i$ & $0.232851-0.244815 i$ & $0.001$ &
        $0.168687-0.434609 i$ & $0.168687-0.434609 i$ & $0.000$ \\
$1.06$ & $0.266277-0.078861 i$ & $0.266278-0.078861 i$ & $0.000$ &
        $0.232499-0.243588 i$ & $0.232518-0.243554 i$ & $0.012$ &
        $0.167607-0.435073 i$ & $0.167911-0.434872 i$ & $0.078$ \\
$1.08$ & $0.266594-0.078406 i$ & $0.266596-0.078407 i$ & $0.001$ &
        $0.232119-0.242404 i$ & $0.232160-0.242413 i$ & $0.013$ &
        $0.166927-0.434789 i$ & $0.166914-0.434523 i$ & $0.057$ \\
$1.10$ & $0.266903-0.077943 i$ & $0.266903-0.077945 i$ & $0.001$ &
        $0.231720-0.241261 i$ & $0.231731-0.241266 i$ & $0.004$ &
        $0.166346-0.435823 i$ & $0.165604-0.435593 i$ & $0.166$ \\
$1.12$ & $0.267204-0.077473 i$ & $0.267204-0.077473 i$ & $0.000$ &
        $0.231303-0.240138 i$ & $0.231303-0.240138 i$ & $0.000$ &
        $0.167251-0.432531 i$ & $0.170116-0.433767 i$ & $0.673$ \\
$1.14$ & $0.267497-0.076995 i$ & $0.267497-0.076995 i$ & $0.000$ &
        $0.230911-0.239018 i$ & $0.230908-0.239019 i$ & $0.001$ &
        $0.163237-0.430724 i$ & $0.166693-0.430621 i$ & $0.751$ \\
$1.16$ & $0.267780-0.076511 i$ & $0.267780-0.076511 i$ & $0.000$ &
        $0.230491-0.237929 i$ & $0.230494-0.237939 i$ & $0.003$ &
        $0.163237-0.427845 i$ & $0.164653-0.429944 i$ & $0.553$ \\
$1.18$ & $0.268054-0.076020 i$ & $0.268055-0.076020 i$ & $0.000$ &
        $0.230020-0.236864 i$ & $0.230022-0.236862 i$ & $0.001$ &
        $0.163013-0.427722 i$ & $0.163300-0.429349 i$ & $0.361$ \\
\hline
\end{tabular}}
\caption{Quasinormal modes of the $\ell=1$ electromagnetic perturbations for the regular black hole II ($M=1$) for three profiles ($n=0,1,2$), calculated using the WKB formula at 16th and 14th orders with Pad\'e approximants. The last column in each block shows the relative difference between the two WKB orders (rounded to three decimals).}
\label{tab:QNM_l1_scalar_n012_merged}
\end{table*}

\begin{table}
\begin{tabular}{c c c c}
\hline
$2 M l^2$ & WKB16 ($m=8$) & WKB14 ($m=7$) & diff.  \\
\hline
$0.02$ & $0.145838-0.771240 i$ & $0.145763-0.771891 i$ & $0.0835\%$\\
$0.1$ & $0.148688-0.763673 i$ & $0.148698-0.763695 i$ & $0.00316\%$\\
$0.2$ & $0.151848-0.752697 i$ & $0.151849-0.752699 i$ & $0.00019\%$\\
$0.3$ & $0.155170-0.741232 i$ & $0.155296-0.741566 i$ & $0.0471\%$\\
$0.4$ & $0.155846-0.730711 i$ & $0.157234-0.730012 i$ & $0.208\%$\\
$0.5$ & $0.152476-0.715703 i$ & $0.155598-0.717208 i$ & $0.474\%$\\
$0.6$ & $0.150139-0.695716 i$ & $0.147812-0.700977 i$ & $0.808\%$\\
$0.7$ & $0.143448-0.679172 i$ & $0.138944-0.679097 i$ & $0.649\%$\\
$0.8$ & $0.129548-0.667670 i$ & $0.126698-0.663298 i$ & $0.767\%$\\
%$0.82$ & $0.124928-0.666869 i$ & $0.123267-0.660977 i$ & $0.902\%$\\
%$0.84$ & $0.119670-0.666102 i$ & $0.119650-0.659022 i$ & $1.05\%$\\
%$0.86$ & $0.114485-0.665600 i$ & $0.116013-0.657512 i$ & $1.22\%$\\
%$0.88$ & $0.109758-0.665900 i$ & $0.112614-0.656555 i$ & $1.45\%$\\
$0.9$ & $0.105752-0.667244 i$ & $0.109797-0.656311 i$ & $1.73\%$\\
%$0.92$ & $0.102803-0.669482 i$ & $0.107901-0.657051 i$ & $1.98\%$\\
%$0.94$ & $0.101155-0.672105 i$ & $0.106975-0.659144 i$ & $2.09\%$\\
%$0.96$ & $0.100662-0.674445 i$ & $0.106480-0.662754 i$ & $1.92\%$\\
%$0.98$ & $0.100720-0.676063 i$ & $0.105495-0.667402 i$ & $1.45\%$\\
$1.$ & $0.100519-0.677042 i$ & $0.103421-0.672081 i$ & $0.840\%$\\
$1.02$ & $0.099375-0.677938 i$ & $0.100452-0.675926 i$ & $0.333\%$\\
$1.04$ & $0.096834-0.679453 i$ & $0.097006-0.679055 i$ & $0.0633\%$\\
$1.06$ & $0.092232-0.682490 i$ & $0.092307-0.682337 i$ & $0.0247\%$\\
$1.08$ & $0.078298-0.692328 i$ & $0.084705-0.685923 i$ & $1.30\%$\\
%$1.1$ & $0.093743-0.629749 i$ & $0.071728-0.687554 i$ & $9.72\%$\\
%$1.12$ & $0.096428-0.660312 i$ & $0.052252-0.677609 i$ & $7.11\%$\\
%$1.14$ & $0.094255-0.664219 i$ & $0.049965-0.640883 i$ & $7.46\%$\\
%$1.16$ & $0.089909-0.663465 i$ & $0.085681-0.626993 i$ & $5.48\%$\\
%$1.18$ & $0.082485-0.657892 i$ & $0.096707-0.628845 i$ & $4.88\%$\\
\hline
\end{tabular}
\caption{Quasinormal modes of the $\ell=1$, $n=3$ electromagnetic perturbations for the regular black hole II ($M=1$)  calculated using the WKB formula at different orders and Pade approximants.}
\end{table}

\begin{table}
\begin{tabular}{c c}
\hline
\hline
 order of expansion & QNM ($n=1$, $\ell=0$, $s=0$) \\
 \hline
 4 & 0.0911838-0.603479 i \\
 8 & 0.0799176-0.580669 i \\
 12 & 0.0757511-0.572925 i \\
 16 & 0.0747444-0.569517 i \\
 20 & 0.0748463-0.568287 i \\
 \text{WKB} & 0.073108 - 0.569709 i \\
 \hline
 \hline
\end{tabular}
\caption{Quasinormal frequencies for scalar perturbations ($n=1$, $\ell=0$) obtained  by the precise Leaver method 
for the truncated metric, shown for successive non-zero orders of the expansion, 
compared with the WKB result obtained for the full black hole metric I. 
All quantities are given in units of the event horizon radius ($r_h=1$); so that $M\approx0.604743$, $4l^4/M^4\approx 9.462833$. For comparison, the Schwarzschild limit in units of the event horizon radius yields 
$\omega_{n=1} = 0.172234 - 0.696105 i$. 
Thus, in the regular black hole case the real part of the oscillation frequency is more than twice smaller than in the Schwarzschild spacetime.
At the same time, the relative deviation between the high-order expansion and the WKB result does not exceed a few percent.
Therefore, the numerical uncertainty of the WKB method remains one to two orders of magnitude smaller than the physical effect observed in the frequency shift.}\label{lasttable}
\end{table}

\textbf{WKB method.} 
The WKB approach relies on connecting approximate solutions constructed in different regions of the effective potential. In particular, one matches the asymptotic solutions, satisfying the quasinormal boundary conditions (purely outgoing waves at infinity and purely incoming at the event horizon), to the local solution obtained from a Taylor expansion of the potential near its maximum \cite{Schutz:1985km,Iyer:1986np,Konoplya:2003ii,Konoplya:2004ip,Matyjasek:2017psv}. 

At the lowest order, the WKB formula reproduces the eikonal approximation, which becomes exact in the limit of large multipole number $\ell \to \infty$. Higher-order corrections usually systematically improve the accuracy by incorporating progressively higher derivatives of the potential at its peak. However, strictly speaking, the WKB series converges only asymptotically and better accuracy at each next WKB order is not always guaranteed. The general WKB expression for the quasinormal frequencies can be written as an expansion around the eikonal limit in the form~\cite{Konoplya:2019hlu}
\begin{small}
\begin{eqnarray}\label{WKBformula-spherical}
\omega^2 &=& V_0 + A_2(\K^2) + A_4(\K^2) + A_6(\K^2) + \ldots \\
&& {} - i \K \sqrt{-2V_2}
\left(1 + A_3(\K^2) + A_5(\K^2) + A_7(\K^2) + \ldots \right), \nonumber
\end{eqnarray}
\end{small}
where the matching condition determining the quasinormal spectrum reads
\begin{equation}
\K = n + \frac{1}{2}, 
\qquad 
n = 0,1,2,\ldots,
\end{equation}
with $n$ being the overtone number.

Here $V_0$ denotes the value of the effective potential at its maximum, while $V_2$ is the second derivative of the potential at that point with respect to the tortoise coordinate. The quantities $A_i$ ($i=2,3,4,\ldots$) represent the $i$-th order WKB corrections beyond the eikonal approximation. Each $A_i$ depends on $\K$ and on derivatives of the effective potential evaluated at the maximum up to order $2i$. 

Explicit expressions for the correction terms were derived at different stages of the method’s development: the second and third orders in~\cite{Iyer:1986np}, the fourth through sixth orders in~\cite{Konoplya:2003ii,Konoplya:2004ip}, the seventh up to thirteenth orders in~\cite{Matyjasek:2017psv} and finally, fourteen to sixteen orders \cite{Matyjasek:2019eeu}. Over the years, this extended WKB framework has been widely applied to the computation of quasinormal modes and grey-body factors in a broad class of black-hole spacetimes (see, e.g., 
\cite{Zinhailo:2019rwd,Fu:2022cul,Xia:2023zlf,Eniceicu:2019npi,Chen:2019dip,Bolokhov:2024ixe,Bolokhov:2024bke,Bolokhov:2023bwm,Bolokhov:2023dxq,Bolokhov:2023ruj,Skvortsova:2025cah,Skvortsova:2024wly,Skvortsova:2024atk,Skvortsova:2023zmj,Konoplya:2020bxa,Lutfuoglu:2025hwh,Lutfuoglu:2025hjy,Lutfuoglu:2025ljm,Lutfuoglu:2025ohb,Lutfuoglu:2025ldc,Lutfuoglu:2025qkt,Lutfuoglu:2026zel,Lutfuoglu:2026xlo,Malik:2025czt,Malik:2025qnr,Zhidenko:2007sj,Zhidenko:2003wq,Zhidenko:2005mv}),
demonstrating its robustness and versatility across different gravitational models. However, when the centrifugal term in the effective potential has a non-standard eikonal form, the WKB method cannot be applied \cite{Konoplya:2017wot}.

\textbf{Time-domain integration.}
An additional and largely independent verification of the WKB results is provided by direct time-domain evolution. Besides comparing different WKB orders and Padé approximants, we integrate the perturbation equation as an initial-value problem and extract the quasinormal frequencies from the resulting ringdown signal. For the numerical evolution we adopt the characteristic Gundlach–Price–Pullin finite-difference scheme \cite{Gundlach:1993tp}, formulated in light-cone coordinates $u=t-r_*$ and $v=t+r_*$. 
The discretization on a null grid with step $\Delta$ takes the form
\begin{eqnarray}
\Psi(N) &=& \Psi(W) + \Psi(E) - \Psi(S) \nonumber \\
&& - \frac{\Delta^2}{8}\, V(S)\left[\Psi(W) + \Psi(E)\right] 
+ {\cal O}(\Delta^4),
\label{Discretization}
\end{eqnarray}
where the grid points are defined as 
$N=(u+\Delta,v+\Delta)$, 
$W=(u+\Delta,v)$, 
$E=(u,v+\Delta)$, 
and $S=(u,v)$. 
This second-order accurate scheme is stable and well suited for tracking the late-time behavior of perturbations.

The dominant frequencies are extracted from the ringdown profile using the Prony method, which fits the time-domain signal by a sum of damped exponentials,
\[
\Psi(t) \approx \sum_{k} C_k e^{-i\omega_k t},
\]
over an appropriate time window where the quasinormal ringing 
dominates. This procedure allows one to determine both the real and 
imaginary parts of $\omega_k$ without prior assumptions about the 
number of contributing modes. The method has been successfully applied 
in numerous studies of black-hole perturbations 
\cite{Malik:2025ava,Malik:2024bmp,Dubinsky:2024gwo,Dubinsky:2024rvf,
Dubinsky:2024hmn,Dubinsky:2024jqi,Qian:2022kaq,Abdalla:2012si,Aneesh:2018hlp,Zhidenko:2008fp,Stuchlik:2025mjj,Konoplya:2025afm}, 
demonstrating reliable agreement with frequency-domain techniques, particularly for the fundamental mode and the first few overtones.

\textbf{Frobenius (Leaver) Method.} For massless perturbations of asymptotically flat black holes, the 
quasinormal spectrum can be computed using the Frobenius expansion 
combined with the continued-fraction technique introduced by Leaver \cite{Leaver:1985ax,Leaver:1986gd}. 

We consider the master perturbation equation expressed in the radial coordinate $r$. This equation is a linear second-order ordinary differential equation possessing regular singular points at the event horizon $r=r_{h}$ and an irregular singular point at spatial infinity $r=\infty$. In order to implement the quasinormal boundary conditions, it is convenient to extract the known asymptotic behavior of the solution.

To this end, we introduce a new function $y(r)$ by writing
\begin{equation}\label{reg}
\Psi(r) = r^{\sigma}  e^{i\omega r} \left(1 - \frac{r_{h}}{r}\right)^{-i\omega / f'(r_{h})} y(r),
\end{equation}
where the power $\sigma$ is fixed by inserting Eq.~\eqref{reg} into the master wave equation and performing an expansion at large $r$ in inverse powers of $r$. With this factorization, the required quasinormal behavior at the horizon and at infinity is already built into the prefactors, so that the function $y(r)$ itself must remain regular at both $r=r_{h}$ and $r=\infty$. Regularity at the horizon allows us to represent $y(r)$ as a Frobenius expansion about $r=r_{h}$ in the form
\begin{equation}\label{Frobenius}
y(r) = \sum_{i=0}^{\infty} a_i \left(1 - \frac{r_{h}}{r}\right)^i.
\end{equation}

Substitution of (\ref{Frobenius}) into the radial equation leads, in the general case, to an $n$-term recurrence relation for $a_n$, which can be reduced to a three-term form by Gaussian elimination. The quasinormal frequencies are then obtained from the corresponding continued-fraction condition ensuring convergence of the series. In our implementation, we employ the integration-through-midpoints procedure developed by Rostworowski \cite{Rostworowski:2006bp}, which was recently applied in numerous works \cite{Stuchlik:2025ezz,Lutfuoglu:2025kqp,Saka:2025xxl}. The convergence of the procedure is improved via the Nollert procedure \cite{Nollert:1993zz,Zhidenko:2006rs}.

\textbf{Quasinormal modes.} The numerical data presented in Tables I–VII reveal a clear qualitative pattern common to both models. 
For the fundamental mode ($n=0$), the dependence of the frequency on the regularization parameter $l$ 
is smooth and relatively weak. Both the real oscillation frequency and the damping rate vary gradually 
as $l$ increases, with no abrupt structural changes. This behavior is consistent with the usual picture 
that the fundamental mode is primarily determined by the geometry near the maximum of the effective 
potential, which itself changes continuously with $l$.

In contrast, the higher overtones exhibit a qualitatively different response. Already for $n=1$ and especially for $n=2,3$, one observes a much stronger sensitivity of the real part of the frequency to the parameter $l$. While the fundamental mode shifts only moderately, each subsequent overtone changes more rapidly as $n$ increases. For sufficiently large $l$, the real part of higher overtones may decrease sharply, demonstrating the characteristic “outburst of overtones.” This effect is particularly visible in Tables III–V, where the deviation of $\mathrm{Re}(\omega)$ for $n=2,3$ becomes significant even though the estimated WKB relative error remains at the level of a few percent. Thus, the observed variation is not a numerical artifact but a genuine physical effect.

An important methodological point is that this overtone outburst can already be detected within the high-order WKB approach. The comparison between the 14th- and 16th-order WKB results with Padé resummation shows excellent agreement for the fundamental mode and first overtone, and still reasonable control for $n=2,3$. Moreover, time-domain integration confirms the same qualitative trend, independently reproducing the fundamental mode (see fig. \ref{fig:TD}). 

However, a fully reliable verification of the overtone frequencies requires a convergent frequency-domain method. For this purpose we employed the Leaver continued-fraction technique \cite{Leaver:1985ax,Leaver:1986gd}. This approach demands that the radial wave equation have coefficients of rational form. To satisfy this requirement, we express all quantities in units of the event horizon radius \textbf{($r_h=1$)}, which fixes the mass parameter as
$
M = \frac{1}{2\sqrt{1 - l^4}}.
$
We then expand the metric function in powers of the regularization 
parameter $l$:
\begin{eqnarray}
f(r) &=& \frac{r-1}{r}
-\frac{l^4 (r^6-1)}{2 r^7}
-\frac{3 l^8 (r^6-1)^2}{8 r^{13}}
\\\nonumber&&-\frac{5 l^{12} (r^6-1)^3}{16 r^{19}}
-\frac{35 l^{16} (r^6-1)^4}{128 r^{25}}
\\\nonumber&&-\frac{63 l^{20} (r^6-1)^5}{256 r^{31}}
+ \mathcal{O}(l^{24}).
\end{eqnarray}

The results obtained with the Leaver method (see Table~\ref{lasttable}) demonstrate that, already at order $\sim l^{20}$, the quasinormal frequencies of the truncated metric converge to the fully precise values with an accuracy well below one percent. The corresponding Leaver frequencies differ from the high-order WKB results by less than two percent.

Importantly, this numerical uncertainty is much smaller than the physical effect under consideration: the variation of the real part of the overtone frequencies reaches tens to more than a hundred percent. Therefore, the discrepancy between methods is negligible compared to the magnitude of the observed overtone outburst. We thus conclude that, for the first few overtones, the high-order WKB method remains reliable in capturing the physical behavior of 
the spectrum. 

The dependence on the parameter $l$ can be understood physically. 
Increasing $l$ enhances the deviation from the Schwarzschild geometry predominantly 
in the near-horizon region, while the asymptotic structure remains unchanged. 
Since higher overtones probe more deeply the near-horizon part of the effective potential, 
their frequencies react more strongly to this modification. 
The fundamental mode, being mainly controlled by the peak region, remains comparatively stable. 
The same qualitative behavior is observed for both models, including the Hayward case 
(the last two tables), indicating that the overtone outburst is not model-specific 
but rather a generic feature of geometrically regular black holes whose deviations 
from Schwarzschild are localized close to the horizon.

\textbf{Conclusions.} Following the general technique, we have constructed a more general solution~\eqref{solution} describing regular, asymptotically flat black holes in the non-polynomial quasi-topological gravity framework introduced in \cite{Borissova:2026wmn}. The procedure is identical to that developed in \cite{Tsuda:2026xjc}. We then analyzed two specific models within this class of black hole solutions.

We have investigated the quasinormal spectrum of scalar, electromagnetic, 
and Dirac perturbations for four-dimensional regular black holes arising 
in non-polynomial quasi-topological gravity. Two representative metric 
profiles were considered, including the case equivalent (after parameter 
redefinition) to the Hayward solution. The analysis was performed mainly 
within the high-order WKB approach with Padé resummation, and cross-checked 
by time-domain integration.

The principal physical result is the clear manifestation of the 
“outburst of overtones.” While the fundamental mode varies smoothly and 
moderately with the regularization parameter, higher overtones 
exhibit progressively stronger sensitivity to deviations from the 
Schwarzschild geometry. Each successive overtone changes more rapidly 
with increasing the regularization parameter, and for sufficiently large deformation the real 
part of the frequency may undergo a pronounced shift. This confirms that 
overtones probe more deeply the near-horizon region, where the geometric 
modifications responsible for regularity are concentrated, whereas the 
fundamental mode is mainly governed by the potential peak region and 
therefore reacts more weakly.

From the methodological point of view, an important observation is that 
this overtone outburst can already be detected within the WKB framework, 
provided sufficiently high orders (14th–16th) with Padé approximants are 
used. The agreement between different WKB orders is excellent for the 
fundamental mode and remains under reasonable control for the first few 
overtones in the parameter range where the effective potential retains 
a standard barrier shape. Time-domain integration (for the fundamental mode) and Leaver method (for overtones) independently confirms the same qualitative behavior. Thus, even though Frobenius-type methods  remain the most precise tool for highly damped modes, high-order WKB 
approximation proves capable of capturing the onset of overtone 
instability in geometrically regular black holes.

Overall, the observed behavior appears to be generic for regular 
black-hole geometries whose deviation from Schwarzschild is localized 
near the event horizon. The enhanced sensitivity of higher overtones 
may therefore serve as a useful probe of near-horizon structure in 
theories where singularities are resolved by higher-curvature effects.

\begin{acknowledgments}
The author thanks A. Zhidenko for useful dicussions. 
\end{acknowledgments}

\bibliography{bibliography}
\end{document}